\documentclass[aps,pre,showpacs,groupedaddress,showkeys,amsmath,amssymb]{revtex4}

\usepackage{graphicx}
\usepackage{bm}
\usepackage{bbm}

\usepackage{epic}
\usepackage{eepic}
\usepackage{pifont}

\usepackage{nicefrac}
\input epsf
\input rotate
\usepackage{graphicx}
\begin{document}
\draft
\title{Lateral migration of a 2D vesicle in unbounded Poiseuille flow}
\author{B. Kaoui (1,2), G. H. Ristow (3), I. Cantat (4),
C. Misbah (1), and W. Zimmermann (3)}
\affiliation{(1) Laboratoire de Spectrom\'etrie Physique, CNRS -
Universit\'e Joseph Fourier / UMR 5588. BP 87, F-38402 Saint-Martin
d'H\`eres Cedex, France\\
(2) Laboratoire de Physique de la Mati\`ere Condens\'ee, Facult\'e
 des Sciences Ben M'Sik,
Casablanca, Morocco\\
(3) Theoretische Physik, Universit\"at Bayreuth,
D-95440 Bayreuth, Germany \\
(4) Groupe Mati\`ere Condens\'ee et Mat\'eriaux, Universit\'e de
Rennes 1, F-35042 Rennes Cedex, France
}%

\email[]{chaouqi.misbah@ujf-grenoble.fr}
\date{\today}

\begin{abstract}
The migration of a suspended vesicle in an unbounded Poiseuille flow
is investigated  numerically in the low Reynolds number limit. We
consider the situation without viscosity contrast between the
interior of the vesicle and the exterior. Using the boundary
integral method  we solve the corresponding hydrodynamic flow
equations and track explicitly the vesicle dynamics in two
dimensions. We find that the interplay between the nonlinear
character of
 the Poiseuille flow and the vesicle deformation causes  a cross-streamline migration of
vesicles towards the center of the Poiseuille flow. This   is in a
marked contrast with a result [L.G. Leal, Ann. Rev. Fluid Mech. {\bf
12}, 435 (1980)] according to which the  droplet moves away from the
center (provided there is no viscosity contrast between the internal
and the external fluids). The migration velocity is found to
 increase  with
the local capillary number (defined by the time scale of the vesicle
relaxation towards its equilibrium shape times the local shear
rate),  but reaches a plateau above a certain value of the capillary
number. This plateau value increases with the curvature of the
parabolic flow profile. We present scaling laws for the migration
velocity.
\end{abstract}

\pacs { {87.16.Dg}
{83.80.Lz}
{87.19.Tt}
}

\maketitle

\section{Introduction}

Vesicles are closed phospholipid membranes suspended in an aqueous
solution. They constitute a first step in a model aiming to capture
elementary ingredients in the study of the dynamics of more complex
entities such as red blood cells. Of particular interest is the
migration of blood cells in the circulatory system. The study of
migration of soft particles under flow presents fundamental (e.g.
understanding the subtle interplay between deformation and the flow)
as well as technological interests (e.g. understanding this problem
may help monitoring vesicles and cell guidance in various
circumstances, such as in microfluidic devices, in the process of
cell sorting-out, and so on...).
\\
In the present work we focus our attention on describing the
dynamics of a single suspended vesicle in  a non-linear shear
gradient of  a plane Poiseuille flow. We consider the small Reynolds
number limit, so that inertia can be neglected. Vesicles in flow
field  have been the subject of extensive studies, both in an
unbounded linear shear flow
\cite{Kraus96,seifert99epje,Misbah2006,beaucourt04a,Gommper,dehaas97,mader06,kantsler06}
as well as in the presence of a wall
\cite{cantat99b,seifert99,sukumaran01,olla,abkarian02}. In an
unbounded  linear shear flow (of low Reynolds number)  a vesicle
does not exhibit a lateral migration with respect to the flow
direction. The presence of a wall breaks the translational symmetry
perpendicular to the flow direction and a vesicle is found to
migrate away from the wall
\cite{cantat99b,seifert99,sukumaran01,olla,abkarian02}.
This so called  lift force is caused by the flow induced
upstream-downstream asymmetry of the vesicle \cite{cantat99b}. More
recently, it has been shown that even a spherical vesicle may
execute a lift force as well, provided that the wall is flexible
\cite{beaucourt04b}. In that case, the wall deformability breaks the
upstream-downstream symmetry.

A nonlinear shear flow has a non translationally invariant shear
rate.  It is therefore  of great interest to understand its possible
contribution  to  a cross-streamline migration process. Here we
focus on the pure bulk effect due to the non-linear shear gradient
of a plane Poiseuille flow alone. Accordingly, we consider a
parabolic flow profile in the absence of bounding walls (in order to
avoid any wall induced lift force), and we consider neutrally
buoyant vesicles so that gravity effect is suppressed.

We find that the curvature of the imposed velocity profile, together
with the vesicle deformability,  causes a systematic migration of a
tank-treading vesicle perpendicularly to the parallel streamlines
towards the flow center-line.  Our results show that the migration
velocity increases with the
 curvature of the flow profile and we provide also scaling results for the migration
velocity as a function of the local capillary number. This behavior
is different from a
 prediction made
 for
droplets\cite{Leal}, according to which droplets should migrate away
from the center of the Poiseuille flow towards the periphery.
Actually, in Ref.\cite{Leal} it is predicted that the direction of
the lateral migration of a droplet depends on the viscosity
contrast. For values between 0.5 and 10 --and particularly in the
absence of a viscosity contrast as treated here--, migration occurs
towards the periphery, while for values smaller than 0.5 or greater
than 10, it occurs towards the center line. We did not observe any
of these scenarios neither from numerical studies.

The scheme of this paper is as follows. In Sec. 2 we present the
model equations and describe briefly the method used to solve the
problem. In section 3 we define the dimensionless parameters and
provide typical experimental values. In Section 4 numerical results
and their discussion are presented. Section 5 is devoted to some
concluding remarks.
\section{Model and method}
\subsection{Hydrodynamical equations and boundary integral method}

The flow of an incompressible Newtonian fluid
 with viscosity $\eta$ and density $\rho$
 is characterized by the dimensionless
Reynolds number,
\begin{equation}
\mbox{Re} = \frac{\rho V_0 R_0}{\eta}, \label{eq: Reynolds}
\end{equation}
where $V_0$ is a characteristic velocity and $R_0$ a characteristic
length of the studied system. In our case we take the size of a
vesicle, which is of the order of $ 10-100\mu$m \cite{Alberts:2001},
as the characteristic length. For such a length and for vesicles
suspended in an aqueous solution subject to shear, with moderate
applied shear rates ($\dot \gamma = V_0/R_0$) that are usually of
the order of $10 s^{-1}$, the Reynolds number is rather small,
$\mbox{Re} \thicksim10^{-2}-10^{-3}\ll1$. Therefore, the flows of
the fluids inside and outside the vesicle, which are taken to be of
the same nature, are well approximated by the Stokes equations,
\begin{eqnarray}
\left\{
\begin{array}{r@{\hskip1ex}c@{\hskip1ex}l}
-{\mathbf\nabla} p+\eta \nabla ^{2}\mathbf{v} &=& \mathbf{f},\\
\nabla \cdot \mathbf{v} &=& 0,
\end{array}
\right.
\label{eq:stokes}
\end{eqnarray}
where $\bf{v}$ is the fluid velocity, $p$ is the pressure, and
$\bf{f}$ is the force imposed by the deformable vesicle membrane on
the two fluids (it is  a local force having a non zero value only at
the membrane). This force is given by the functional derivative of
the vesicle membrane energy with respect to the membrane
displacement, as discussed in the next subsection.

Using the boundary integral method
\cite{Ladyzhenskaya69,pozrikidis92} we solve the equations
(\ref{eq:stokes}) in  two-dimensional space. The velocity field at
any point in the fluid (or at the membrane; the membrane velocity is
equal to that of the adjacent fluids provided that the membrane is
not permeable, and that there is no slip at the membrane)
can be  written as a superposition (due to linearity of the Stokes
equations) of two terms, namely the contribution from the vesicle
boundary, plus a contribution due to  the undisturbed applied
Poiseuille flow $\bf{v}(\mathbf{r})_{\text{Pois}}$ (to be written
below),
\begin{equation}
v_{i}(\mathbf{r})=\int_{\partial \Omega } d\mathbf{r}^{\prime
}\mathbf{G}_{ij}(\mathbf{r}-\mathbf{r}^{\prime
})f_{j}(\mathbf{r}^{\prime}) ds(\mathbf{r}^{\prime})+
v_{i}(\mathbf{r})_{\text{Pois}}. \label{eq: vel}
\end{equation}
Here $\partial \Omega$ refers to the vesicle boundary.
$\mathbf{G}_{ij}$ denotes the Oseen tensor, also called Green's
function of the Stokes equations. Since we focus on  dynamics of a
vesicle suspended in an unbounded domain, we use the two-dimensional
free space Green's function, that has the following expression
\cite{pozrikidis92,cantat03}:
\begin{equation}
\mathbf{G}_{ij}=\frac{1}{4\pi \eta }\left( -\delta _{ij}\ln r+\frac{%
r_{i}r_{j}}{r^{2}}\right),
\end{equation}
where $r\equiv \left\vert \mathbf{r}-\mathbf{r}^{\prime }\right\vert
$ and $r_{i}$ is the $i^{th}$ component of the vector
$\mathbf{r}-\mathbf{r}^{\prime }$.

Equation (\ref{eq: vel}) is valid in the fluid as well as at the
membrane. In order to obtain the membrane velocity we replace
${\mathbf r}$ by the membrane vector position. Numerically, the
vesicle membrane contour (in 2D) is discretized, as explained in
Ref.~\cite{cantat03}. After evaluating the membrane force which
enters the right hand side of Eq. (\ref{eq: vel}) (see next section
for the force evaluation), the velocity is then evaluated at each
discretization point using Eq. (\ref{eq: vel}). The displacement in
the course of time of the vesicle membrane is obtained by updating
the discretization points after each time iteration, using an Euler
scheme, $\mathbf{r}(t+dt)=\mathbf{v}(\mathbf{r},t)dt+\mathbf{r}(t)$.
In the following section we shall discuss  in more detail
 the forces and the constraints.

\subsection{Membrane force and comparison with droplets}

The vesicle membrane is a bilayer made of phospholipid molecules
having a hydrophilic head and two hydrophobic tails. At room
temperature (and at physiological temperature as well) the membrane
is fluid. The membrane  can be viewed as  a two-dimensional
incompressible fluid. This incompressibility property implies the
inextensibility of the membrane, and therefore, the conservation of
local area. Moreover, since the vesicle encloses an incompressible
fluid and the membrane permeability  is very small, the vesicle
volume must be  a conserved quantity. Due to membrane
impermeability, the membrane velocity is equal to the fluid velocity
of the adjacent layer. Consequently, and because we use explicitly
the incompressibility condition  $\nabla .{\mathbf v}=0$ for fluids,
the enclosed volume is automatically conserved. The area of the
membrane is not conserved automatically (think of a droplet that can
spread out on a substrate; its volume is conserved while its surface
increases). Thus, in order to fulfill membrane local area (or
perimeter in 2D) conservation, one must introduce a surface (local)
Lagrange multiplier. This Lagrange multiplier  $\zeta(s,t)$ depends
on the curvilinear coordinate $s$ along the vesicle contour and on
time, since the incompressibility should be fulfilled locally and at
each time. $\zeta$ is the surface analog of the pressure field
$p({\mathbf r, t})$ which enforces local volume conservation of  a
3D fluid.

Due to the fact that  phospholipidic molecules are bound to the
membrane (there is no exchange between the bilayer and the
surrounding solution), the local area (or area per molecule) remains
constant in the course of time. This is a major difference with
droplets, where the surface molecules can easily migrate into the
underlying bulk and vice versa, causing thus a change of area. One
may say that the surface molecules of a droplet are in contact with
a large reservoir of molecules in the bulk (the reservoir fixes the
chemical potential, while the number of molecules at the surface is
a fluctuating quantity). For a droplet, the question of energy per
unit surface makes  sense: it corresponds to the increase in energy
due to the transfer of a molecule from the bulk towards the
surface. This is surface energy, or surface tension (note that for a
liquid the surface energy and surface tension refer to the same
quantity, while this is not the case for a solid). The surface force
for a droplet is given by the classical Laplace law
\begin{equation}
{\mathbf f_d}= -\sigma H {\mathbf n} \label{surfacetension}
\end{equation}
where $\sigma$ is the surface tension, ${\mathbf n}$ is the outward
normal unit vector, and $H$ is the surface mean curvature.  Note,
that in the 2D case, as treated here, there is only one curvature
$H$ and  $H$ is by convention counted to be positive for a circle.

As a  vesicle membrane does not naturally change its area, the
notion of cost of energy per unit area can not be evoked. Moreover,
one may think of changing the area by applying  an external force
and  the surface energy  is in this case the work associated with
the applied force. This notion is, however, quite different from
surface energy of a droplet. Here, for vesicles, we must rather
refer to  a surface stress, since by applying a force the
intermolecular distance (due to stretching, for example) is
modified. In contrast, a droplet may change its area without
affecting the intermolecular  distance. If we do not apply a large
force in order
 to stretch the microstructure of the
lipid layer, then membrane incompressibility is fulfilled.
Hydrodynamical forces (as those encountered in this problem) are too
small in comparison with cohesive forces, so that membrane
incompressibility is safely satisfied.

From the mechanical point of view, the membrane can be viewed as a
thin plate, where the soft (or easy) mode is the bending one. The
corresponding energy is given by the Helfrich curvature energy
\cite{Helfrich1973}. This reads in 2D
\begin{equation}
E_C=\frac{\kappa}{2}\int_{\partial \Omega}H^{2}ds,
\end{equation}
where $\kappa$ is the membrane rigidity and $H$ the local membrane
curvature. $ds$ is the elementary arclength along the vesicle
contour. Note that for the sake of simplicity, we do not account for
a spontaneous curvature (a constant spontaneous curvature, $H_0$,
may be included by substituting $H$ by $H-H_0$). In order to take
into account the area (perimeter in 2D) constraint, we must add to
the above energy the following contribution $\int_{\partial
\Omega}\zeta(s,t) ds$, so that the total energy reads
\begin{equation}
E=\frac{\kappa}{2}\int_{\partial \Omega}H^{2}ds + \int_{\partial
\Omega}\zeta(s,t) ds \label{eq:energy}
\end{equation}
where $\zeta(s,t)$ is a local Lagrange multiplier. Note, that global
conservation of the perimeter would be unphysical, because  it would
allow  at some range of the membrane  an arbitrarily large
stretching and at the same time at another point a corresponding
compression in a way that the global length is preserved. As
discussed above, stretching or compression is possible only under
the
 action of strong forces, of the
order of cohesive forces.

The force acting on the membrane is obtained from the functional
derivative of the vesicle energy $E$ with respect to a membrane
displacement. The resulting force has been already used
 previously (e.g. \cite{cantat99b,cantat03}), but a derivation has not  been
presented. A detailed derivation is given in the appendix. The
resulting force is
\begin{equation}
\mathbf f = \left[ \kappa\left( \frac{\partial^2 H}{\partial s^2} +
\frac{{H}^3}{2}\right) - H \zeta \right]\mathbf{n} +
\frac{\partial\zeta}{\partial s}\mathbf{t}\ , \label{eq: f_mem}
\end{equation}
where $\mathbf{t}$ is the tangent vector (and recall that
$\mathbf{n}$) is the normal vector). This force is composed of a
normal as well as a tangential contribution. If $\zeta$ is constant,
then only the normal part survives because of the following reason.
If $\zeta$ is constant, the tension-like force (which is a vector)
associated with $\zeta$ is tangential to the curve, and has the same
magnitude at both extremities of an arc element $ds$ (which can be
taken to be a portion of a circle, provided that $ds$ is small
enough). It follows, that the sum of the two forces is directed in
the normal direction. If, on the contrary, $\zeta$ changes along the
contour, then the two values at the extremities of $ds$ are
different, and the force has, besides a normal part, a tangential
one, which is given by $({\partial\zeta}/{\partial s})\mathbf{t}$.
On the other hand, the bending energy depends on  the curvature
(which is a geometrical quantity). It follows that the only force
that is able to change the shape of a geometrical surface (i.e. a
mathematical boundary having no internal physical structure) must be
normal to the surface. Finally, note that the term $-\zeta H
\mathbf{n}$ has the same structure as the force due to surface
tension of a droplet Eq.~(\ref{surfacetension}). There is, however,
a significant physical difference: for a droplet $\sigma$ is an
intrinsic quantity
 which represents the cost in energy for moving a  molecule from
the bulk (surrounded by other molecules) to the surface (and thus it
looses some neighbors). In the present problem $\zeta$ is a Lagrange
multiplier which must be determined a posteriori by requiring a
constant local area. $\zeta$ is not an intrinsic quantity, but
rather it depends on other parameters (like $\kappa$, the vesicle
radius, etc...).

\subsection{Fulfilling local membrane area}

In principle, from Eq.(\ref{eq: vel}) we can determine the membrane
velocity, if the force and the initial shape are given. The force
(\ref{eq: f_mem}) contains geometrical quantities (like the normal
and $H$) which are determined from the initial shape, plus a
function $\zeta(s,t)$, which is unknown a priori. Numerically, the
following method has been tested. An initial shape (typical an
ellipse) and an initial $\zeta$ (typical constant along the contour)
have been chosen. Then the geometrical quantities appearing in the
force can be calculated (the method of discretization of the
integral equation (\ref{eq: vel}) has been discussed in
\cite{cantat03}). This allows one to evaluate the right hand side of
(\ref{eq: vel}) at initial time. The membrane velocity at this time
is thus fixed.  We then displace each membrane element according to
the computed velocity, and by this way we obtain a new shape.
However, the new shape does not fulfill, in general, the local
membrane incompressibility. A local stretching (or compression) of
the membrane takes place as long as the projected divergence of the
velocity field of the fluid adjacent to the membrane is non zero. We
thus must adjust the appropriate function $\zeta(s)$ in order to
fulfill this condition. The condition that the projected divergence
must vanish reads
\begin{equation}
({\mathbf I}- {\mathbf n} {\mathbf n}):{\nabla} {\mathbf v} =0
\end{equation}
where $\mathbf I$ is the identity tensor, and ${\mathbf n}{\mathbf
n}$ stands for the tensor product (${\mathbf I}- {\mathbf n}
{\mathbf n}$ is the projector on the contour). The above relation
can be viewed as an implicit equation for $\zeta(s)$, similar to
$\nabla.{\mathbf v}=0$ which fixes the pressure field in 3D fluids.
This way of reasoning is quite practical in the analytical study of
vesicles \cite{Misbah2006}. From the numerical point of view, this
way has suffered from several numerical instabilities. We have thus
introduced another approach \cite{cantat03} as outlined below.

In a 2D simulation, when discretizing the vesicle membrane contour,
the vesicle perimeter conservation constraint could be achieved
without dealing with the local Lagrange multiplier entering the
membrane force given  by Eq.~(\ref{eq: f_mem}). This constraint
could be fulfilled in another  and  more convenient way. For that
purpose we have used  a straightforward method based on the fact
that two material representative points on the membrane are attached
to each other by  strong cohesive forces which we describe by
quasi-rigid springs, so that we can achieve in numerical studies
less than $1 \%$  variation of the area. By this way an additional
parameter $k_{s}$ is introduced, which is the spring constant
\cite{cantat03}, $\zeta ^{N}(i)=k_{s}(ds(i)-ds^{0}(i))$. By choosing
$k_{s}$ large enough (in order to keep the membrane
quasi-incompressible) the discretization step $ds(i)$ is kept as
close as possible to its initial value $ds^{0}(i)$. Typically in
units where $\eta=\kappa=1$ and where the typical radius of vesicles
is of order unity, a value of $k_s=10^{3}$ has proven to be
sufficient.

\subsection{Applied flow}

The applied Poiseuille flow $\bf{v}(\mathbf{r})_{\text{Pois}}$
has the  following form
\begin{eqnarray}
\left\{
\begin{array}{r@{\hskip1ex}c@{\hskip1ex}l}
v_{\text{x}}(\mathbf{r})_{\text{Pois}} &=& v_{\text{max}}\left[ 1-\left( \frac{y}{w}\right) ^2\right] , \\
v_{\text{y}}(\mathbf{r})_{\text{Pois}} &=& 0,
\end{array}
\right.
\label{eq: poiseuille}
\end{eqnarray}
where $v_{\text{max}}$ is the maximum velocity at the centerline
located at $y=0$ and $2w$ is the width of the Poiseuille profile.
For our simulations we choose always the aspect ratio $R_0/w \ll 1$
in order to keep $v_{\text{x}}(w)=v_{\text{x}}(-w)=0$ practically
unperturbed by the presence of the vesicle.
\section{Dimensionless numbers}
It is convenient to use in the simulation a dimensionless parameter
that we call the local capillary number, which we define as
\begin{equation}
C_a(\mathbf{r})=\tau \dot{\gamma}(\mathbf{r}) \,.\label{eq: ca}
\end{equation}
$\tau$ is the characteristic time for a vesicle to relax to its
equilibrium shape (in the absence of imposed flow), which is given
by
\begin{equation}
\tau = \frac{\eta R_{0}^3}{\kappa}\,, \label{eq: tau}
\end{equation}
$\dot{\gamma}(\mathbf{r})$ is the local shear rate of the applied
Poiseuille flow, that can be evaluated from the corresponding
velocity profile
\begin{equation}
\dot{\gamma}(\mathbf{r})=\dfrac{\partial
v_{\text{x}}(\mathbf{r})}{\partial y}= -\left(
2\dfrac{v_{\text{max}}}{w^2}\right) y=cy.
\end{equation}
Here $c$ is the curvature of the Poiseuille flow  profile, which is
 given by $c=\partial^2 v_x/\partial^2 y$. In the numerical scheme,
 there is another capillary number associated with the tension
$k_s$ (or spring constant), and is defined by $C_{as}={\eta
R_{0}}/{k_s}$. In most simulations we have kept  the ratio
$C_a/C_{as}$ small (of the order of $10^{-3}$). This means that the
time scale for stretching/compression of the membrane is fast in
comparison to bending. In other words, local area conservation is
adiabatically slaved to the overall shape evolution.

As a  characteristic velocity we choose,
\begin{equation}
V_0={R_0\over \tau}={\kappa\over \eta R_0^2}~,
\end{equation}
with  $R_0\equiv L/2\pi$, where $L$ is the vesicle perimeter.
Hereafter we shall use $\tau$ as a unit of time, $R_0$ of length and
$V_0$ as unit of velocity. For typical experimental values of $\eta$
(e.g. water), and by using standard values for vesicles $\kappa\sim
20 k_B T$ ($k_BT$ is the elementary thermal excitation energy) and
$R_0\sim 10\mu m$ one finds $\tau\sim 10 s$ and $V_0\sim 1\mu m/s$.
In the following (and especially in the figures of the simulation),
when a velocity is written in terms of a number (without units) this
means it is expressed in unit of $V_0$. Since $V_0$ is typically of
order $1\mu m /s$, the velocity is given practically in $\mu m/s$.
The reported values for the velocities in experiments on vesicles
are in the range of $1-100\mu m/s$ \cite{vitkova04,mader06}. In the
circulatory system, data are  for the shear stress at the vessel
wall are well documented\cite{fung}.  For example,
 in arteries\cite{fung} the shear stress is of about
$1-2\; Pa$. Dividing this by the Plasma viscosity (close to that of
water), one finds the shear rate at the wall, $\dot\gamma_{wall}\sim
10^{3} s^{-1}$. The velocity at the center of the arteries is of
about $\dot\gamma_{wall} w$. For  small arteries $w\sim 100\; \mu
m$, so that $v_{max}\sim 10^{5}\mu m/s$ (for venules, one has about
$v_{max}\sim 10^{4}\mu m/s$). The chosen values in the simulations
(see figures in the next section) are rather in the experimental
range for vesicles, but are not far away from data on blood flow in
arteries.

\section{Simulation results and discussion}\label{sec: Poiseuille}

\begin{figure}
\includegraphics[width=5cm]{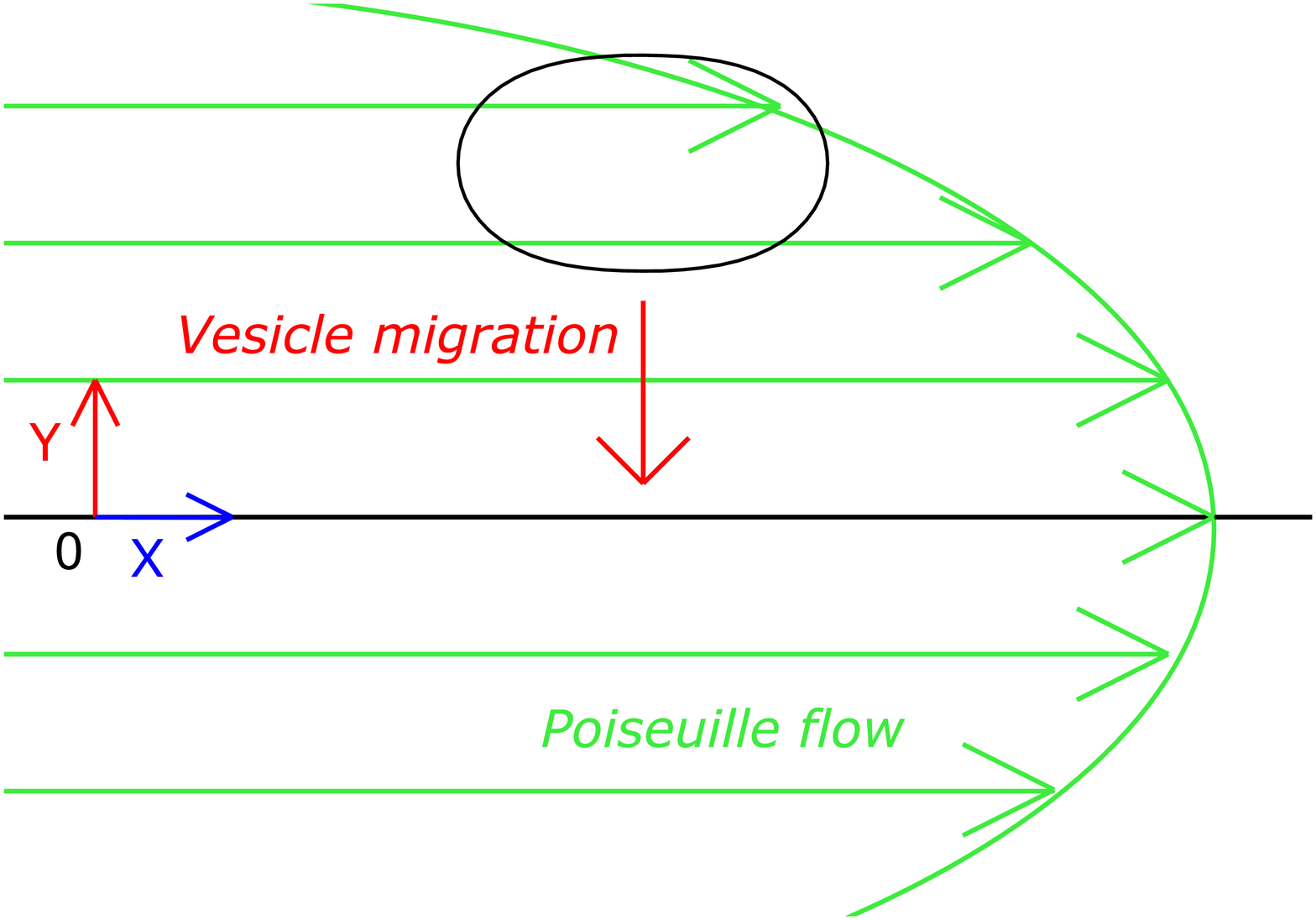}
\includegraphics[width=5cm]{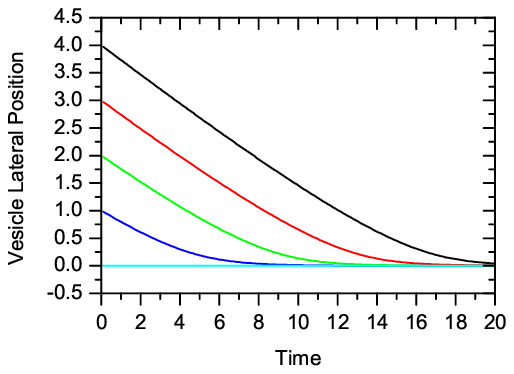}
\caption{\label{Figure1}%
(Color) The left plot  shows a schematic representation of the migration of
a vesicle in an parabolic flow profile corresponding to an unbounded
plane Poiseuille flow. The right plot shows the time evolution of
the lateral position of a vesicle released initially at five
different initial positions ${y_0=0,1,2,3,4}$ in units of the
characteristic time $\tau$ (right). }
\end{figure}
The left part of Fig.~\ref{Figure1} illustrates
 a free  vesicle in an  unbounded plane Poiseuille flow, the dynamics of which
is investigated. The right part of Fig.~\ref{Figure1} shows the time
evolution of the lateral position of a vesicle which has been
released initially from five different vertical positions:
${y_0=0,1,2,3,4}$. In most cases we have studied quasi-circular
vesicles placed in a Poiseuille flow characterized by
$v_{\text{max}}=800$ and $w=10$. In all situations treated so far,
vesicles migrate to the center of the Poiseuille flow where no
further lateral migration is observed. The position gives the
distance from the center of the Poiseuille flow measured in units of
the vesicle effective radius (as defined in the last section). All
the curves are linear in a large range and deviations from this
linear law occur only close to the center of the Poiseuille flow, in
a range smaller than the vesicle size.
\\
\begin{figure}
\centerline{\includegraphics[width=16cm]{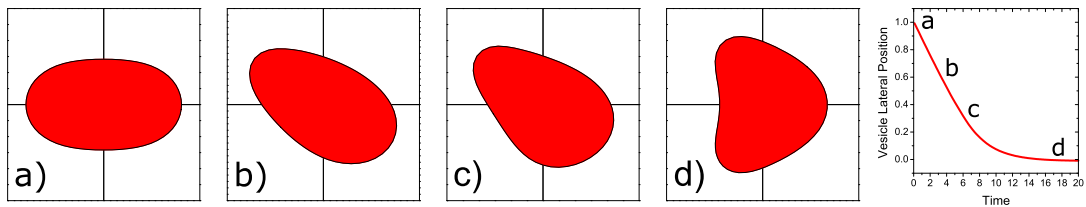}}
\caption{\label{Figure2}%
(Color) The shape of the vesicle changes from an initially elliptical shape in part
a) to the final parachute shape in d) when it migrates towards
Poiseuille flow center-line, reduced area $\nu = 0.90$.}
\end{figure}
During the migration the vesicle shape undergoes deformations due to
the hydrodynamic stresses imposed by the Poiseuille flow on the
 membrane. The vesicle is deformed and tilted
until reaching a quasistationary
orientation which is oblique with respect to
 the parallel streamlines. Figure~\ref{Figure2} shows different vesicle shapes
deformation occurring in our simulations during the migration, from
an initially elliptical shape
at the
initial position $y_0=1$ shown in Fig.~\ref{Figure2}(a), to a final
parachute shape at the center of the Poiseuille flow as shown in
Fig.~\ref{Figure2}(d). More or less similar  parachute shapes are
known for capsules and red blood cells \cite{barthes-biesel,Secomb} as they have been observed
also experimentally for vesicles in Ref.~\cite{vitkova04},
but  all these  examples concern capillary flows. Thus, the present
result shows that the parachute is not necessarily enforced by a
wall, but rather by the curvature of the Poiseuille flow itself.
\begin{figure}
\epsfxsize=0.3\linewidth
 \includegraphics[width=5cm]{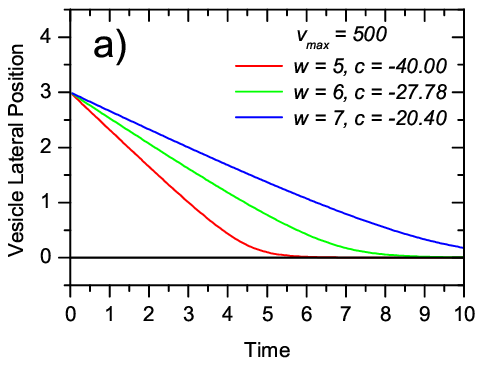}
 \epsfxsize=0.3\linewidth
  \includegraphics[width=5cm]{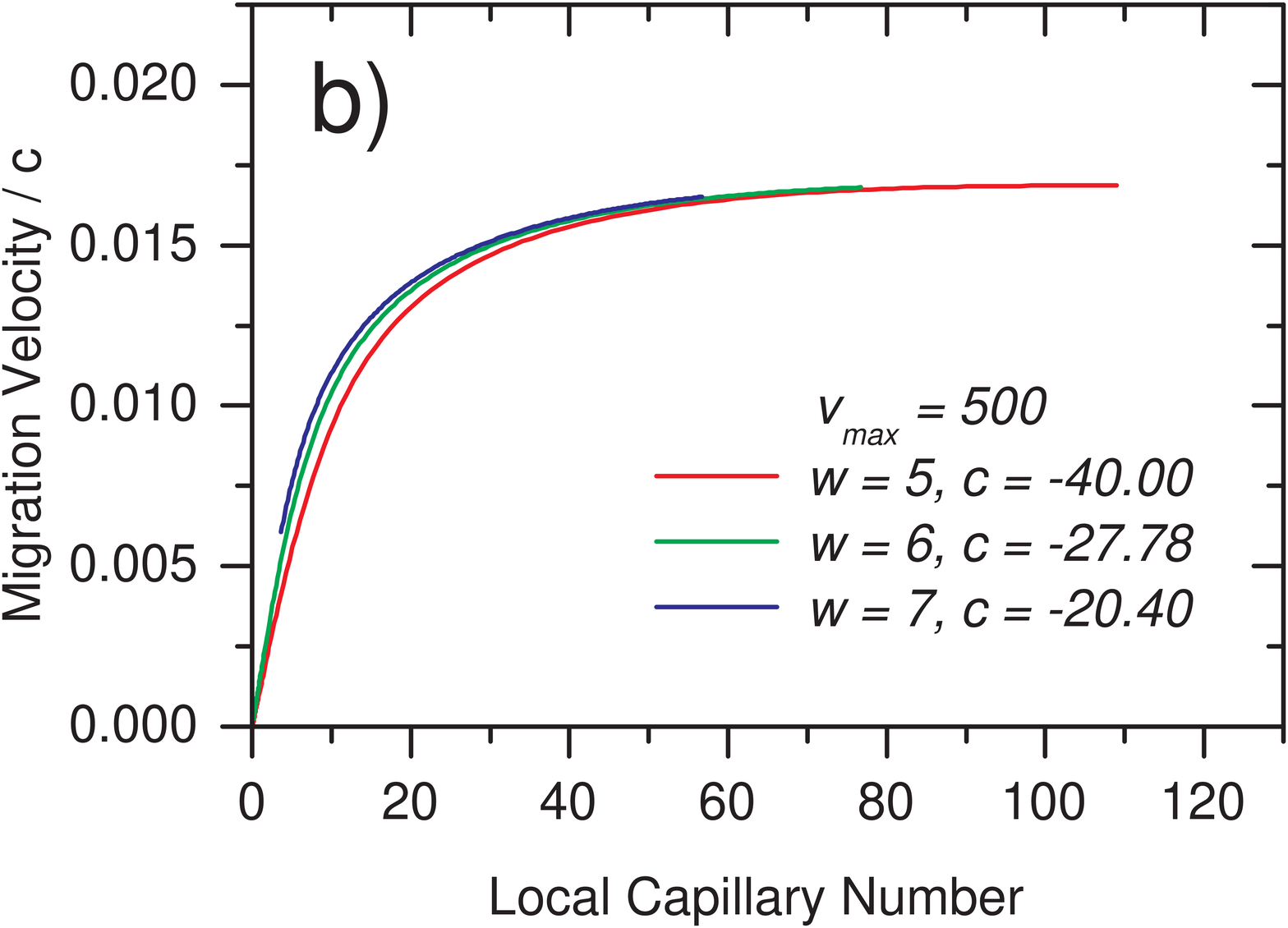}
  \epsfxsize=0.3\linewidth
\includegraphics[width=5cm]{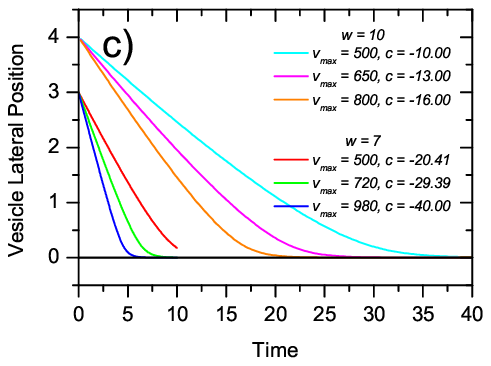}
\epsfxsize=0.3\linewidth
 \includegraphics[width=5cm]{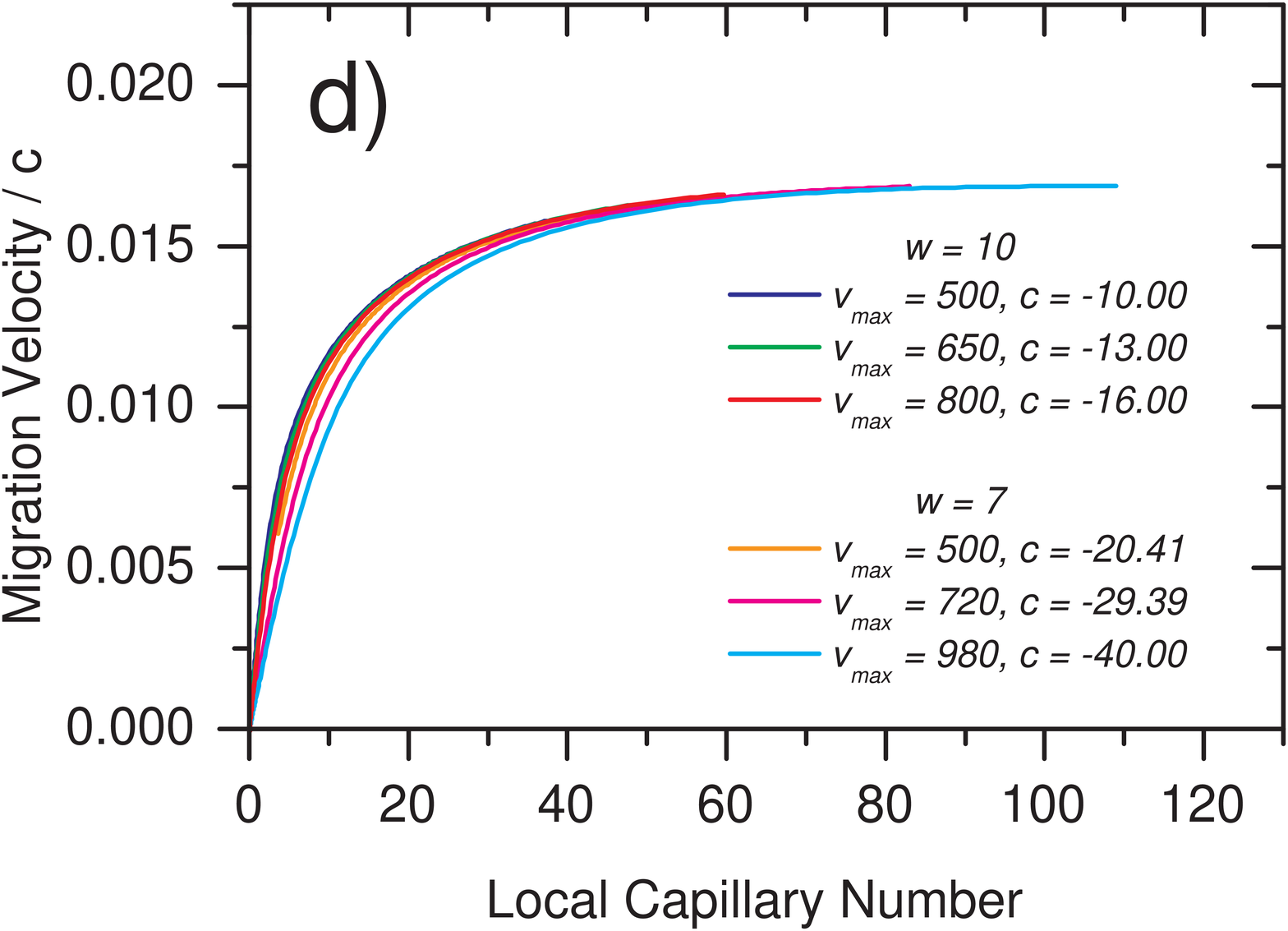}
 \epsfxsize=0.3\linewidth
  \includegraphics[width=5cm]{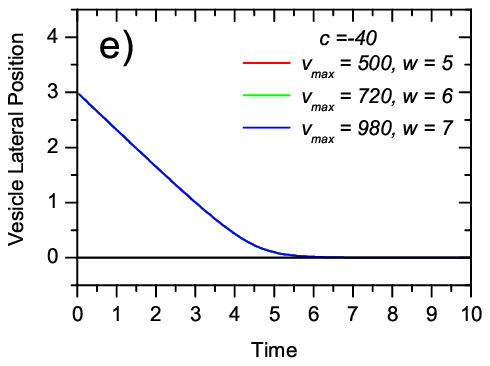}
  \epsfxsize=0.3\linewidth
\includegraphics[width=5cm]{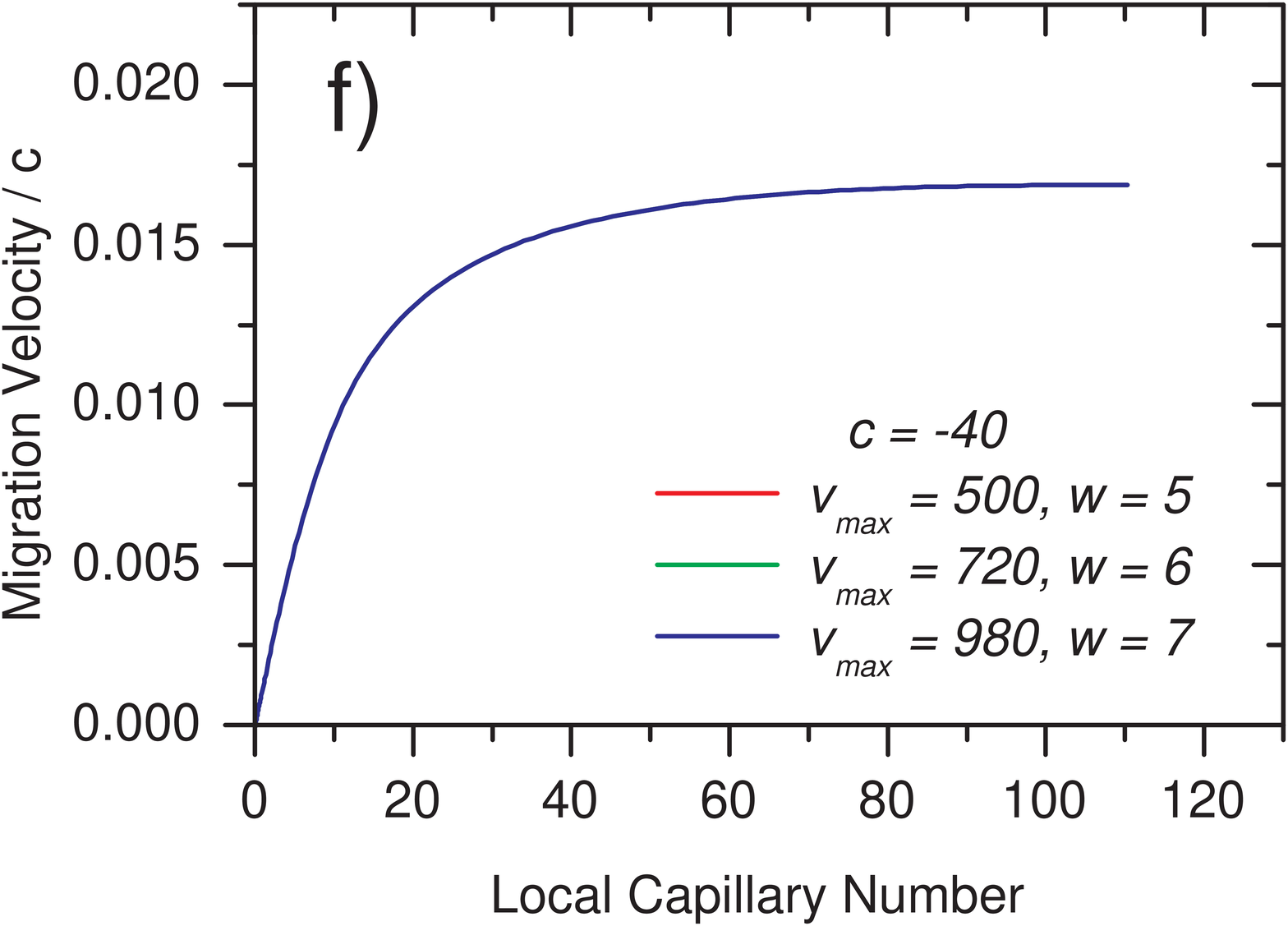}
\caption{\label{Figure3}%
(Color) Time evolution of the  vesicle position in an unbounded
Poiseuille flow and its corresponding normalized migration velocity
versus the local capillary number for different values of
$v_{\text{max}}$, $w$ and $c$ (see text). The data correspond to
situation where initial transients have  decayed.}
\end{figure}

Before the vesicle reaches the center of the Poiseuille flow, it
acquires an asymmetric shape as depicted in Fig.~\ref{Figure2}(b)
and in Fig.~\ref{Figure2}(c). This asymmetry, which is caused by the
non-uniform shear rate across the vesicle,  is crucial   for
cross-stream line migration of vesicles in a plane Poiseuille flow.

As stated above,  a drop (having no viscosity contrast with the
ambient fluid) is predicted to drift towards the periphery
\cite{Leal}. Thus, vesicles and droplets behave quite differently.
In section IIB we have presented the main differences between
vesicles and droplets, both from the physical and the mathematical
point of view. Which of these differences, albeit very important,
may explain the differences in the migration direction, is not clear
at present. For vesicles, we have explored a large domain of
parameter space and  in all cases the vesicle migrate towards the
center.

The migration velocity depends on various parameters. Of particular
importance are  the curvature of the velocity profile of the
Poiseuille flow and  the local capillary number, as discussed in the
following. To the best of our knowledge,  there is in the absence of
a wall  no lateral migration in a linear shear flow. In the presence
of a flow !!! with a nonlinear shear gradient, migration becomes
possible, provided that the shear rate changes on the scale of the
vesicle size.
 Therefore,  curvature of the Poiseuille flow profile plays an
essential role, but more precisely,  the magnitude of the local
capillary number, which determines essentially the vesicle
deformation (which loses the up-down symmetry due to the shear
gradient), is the most relevant quantity.

The dependence of the migration velocity on the
local capillary  is
shown in Fig.~\ref{Figure3} for different values of
$v_{\text{max}}$, $w$ and $c$, after the decay of an initial transient.
In Fig.~\ref{Figure3}(a) and in Fig.~\ref{Figure3}(b) we kept the
value of $v_{\text{max}}$ fixed and we investigated  the vesicle
migration by varying the value of $w$. For smaller values of $w$,
which corresponds to larger values of the curvature $c$, the vesicle
migrates faster towards the center of the Poiseuille flow.
Fig.~\ref{Figure3}(b) shows the data collapse by plotting the
migration velocity normalized to the curvature versus the local
capillary number. In Fig.~\ref{Figure3}(c) and in
Fig.~\ref{Figure3}(d) we kept $w$ fixed and we examined the effect
of varying the value of $v_{\text{max}}$ for each value of $w$. The
vesicle migrates faster with increasing values of  $v_{\text{max}}$
for every fixed $w$, because the curvature $c$ increases with
$v_{\text{max}}$. Data collapse is again obtained in
Fig.~\ref{Figure3}(d) by plotting the normalized migration velocity
versus the local capillary number. The data collapse is more
pronounced for smaller  values of the curvature. In
Fig.~\ref{Figure3}(e) and Fig.~\ref{Figure3}(f)  we have varied
$v_{\text{max}}$ and $w$ in such a way to keep the curvature fixed.
We find that the vesicle migrates in this case to the Poiseuille
flow center-line for the three parameters combinations exactly (i.e.
quantitatively the same results) in the same manner, which
emphasizes again the important role of the nonlinear shear field.
From the above study, we can conclude that   the migration velocity
in an unbounded Poiseuille flow normalized to the curvature $c$
should be described by the following universal scaling law:
\begin{eqnarray}
\dfrac{v_{\text{migration}}(y)}{c}\sim f\left[ C_a(y)\right]\,.
\label{eq:law}
\end{eqnarray}
The extraction of this law  is  based on results of Figs.
\ref{Figure3}b, \ref{Figure3}d, and \ref{Figure3}f. The function $f$
is universal and depends only on $C_a$. The analytical form of the
universal function is not at present.  A first step towards this
issue is to develop an analytical theory  in the  small deformation
limit, as in Ref.\cite{Misbah2006}. The small deformation limit
provides us with nonlinear differential equations for the shapes and
the migration of the vesicle instead of the less tractable
integro-differential equation (\ref{eq: vel}). With this
semi-analytical approach progress  seems  more likely for  both, for
an understanding of the migration direction, and the determination
of the scaling function $f$.

If the initial vesicle shape is not quasi-circular but elliptical,
we find a similar behavior as depicted in Fig.~\ref{Figure3}.
The deformability of the vesicle, which depends on the bending
rigidity $\kappa$, is a further
ingredient for migration. 
By increasing the
bending rigidity $\kappa$ the local capillary number
 $C_a \propto \kappa^{-1}$ decreases and so does the
migration velocity. This leads also to the conclusion
that a rigid particle, corresponding to very large values of
$\kappa$, will not exhibit a lateral migration in parabolic shear
flow in the Stokes limit. It has been shown earlier that
rigid spheres migrate only  due to the
contribution of $(v \cdot \nabla)v$ in the Navier-Stokes equation
 \cite{Joseph:94.1},
which is beyond the Stokes limit.
 Similar trends as for a vesicle
 are obtained for deformable bead-spring
models~\cite{leonhard00}, where
the migration velocity decreases
 with increasing rigidity of the
tumbling  object, corresponding also to increasing values of the
spring constant. Indeed the vesicle deformability is besides the
nonlinear shear gradient the main ingredient for the  lateral
migration in the Stokes limit. A vesicle in unbounded Poiseuille
flow undergoes large deformations ($C_a\gg1$, see Fig.~\ref{Figure3}
) caused only and mainly by the curvature  of the velocity profile.

\section{Conclusions}
The dynamical behavior of a single vesicle placed in an unbounded
plane Poiseuille flow has been investigated numerically. We found
that the vesicle migrates during its tank-treading motion towards
the center of a parabolic flow profile. The migration velocity is
found to increase with the local capillary number (defined by the
time scale of the vesicle relaxation towards its equilibrium shape
times the local shear rate), but reaches a plateau above a certain
value of the capillary number. This plateau value increases with the
curvature of the parabolic flow profile $c$. When the vesicle
reaches this final equilibrium position, its lateral migration
velocity vanishes and it continues to move with a  parachute shape
parallel to  the flow direction.  We found that the migration
velocity normalized to the curvature $v_{\text{migration}}/c$
follows essentially a  universal law where the universal function
depends on the local capillary number $C_a$, namely
$v_{\text{migration}}/c\sim f(C_a)$. A droplet having no viscosity
contrast seems to move away from the center \cite{Leal}, which is in
a marked contrast to vesicles, for which we found migration towards
the center. This difference is not yet fully understood, and is
currently under investigation. In forthcoming publications our
calculations will be also extended to droplets in order to extract
the main source of difference between the vesicle and drop
migration.

\acknowledgements We would like to thank A. Arend and S. Schuler for
enlightening and very helpful discussions. C. M. acknowledges a
financial support from CNES (Centre National d'Etudes Spatiales) and
from CNRS (ACI Mod\'elisation de la cellule et du myocarde). W. Z.
acknowledges financial support from DFG (German science foundation)
via the priority program SPP 1164. B.K. and C.M.  acknowledge a
Moroccan-French cooperation programme (PAI Volubilis).
\appendix*
\section{Derivation of the membrane force}
In a  two spatial dimension  the vesicle membrane is represented by
a one-dimensional closed contour. The corresponding membrane energy
is an integral  over this contour,
\begin{equation}
E=\frac{\kappa}{2}\int_{0}^{L}H^{2}(\mathbf{r})ds(\mathbf{r})+
\int_{0}^{L}\zeta(\mathbf{r})ds(\mathbf{r}), \label{eq:energyapp}
\end{equation}
where $L$ is the vesicle perimeter (i.e. the length of the contour)
and $\mathbf{r}$ the membrane vector position. Let, \begin{equation}
E_{C}=\frac{\kappa}{2}\int_{0}^{L}H^{2}(\mathbf{r})ds(\mathbf{r}),
\label{eq:energy_c}
\end{equation}
and,
\begin{equation}
E_{T}=\int_{0}^{L}\zeta(\mathbf{r})ds(\mathbf{r}),
\label{eq:energy_t}
\end{equation}
\begin{figure}
\includegraphics[width=5cm]{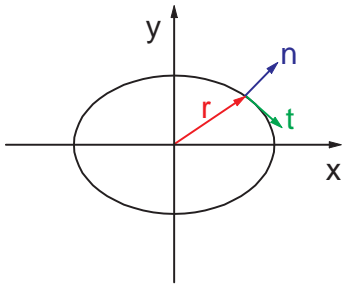}
\centering
\caption{\label{appendix}%
(Color) A schematic showing the vector position $\mathbf{r}$, the normal  $\mathbf{n}$ and
the tangent  $\mathbf{t}$ unit vectors.}
\label{fig:vectors}
\end{figure}
 The counterclockwise
tangent unit vector (see Fig. \ref{fig:vectors}) is given by,
\begin{equation}
\mathbf{ t}=\frac{\partial \mathbf{r}}{\partial s},
\label{eq:tangent}
\end{equation}
and its derivative with respect to $s$ defines the curvature,
\begin{equation}
\frac{\partial \mathbf{ t}}{\partial s}=-H\mathbf{ n},
\label{eq:der_tangent}
\end{equation}
where $\mathbf{ n}$ is the outward unit vector normal to the curve.
The derivative of $\mathbf{ n}$ with respect to $s$ gives,
\begin{equation}
\frac{\partial \mathbf{ n}}{\partial s}=H\mathbf{ t}.
\label{eq:der_normal}
\end{equation}
Using Eq.~(\ref{eq:tangent}) and  Eq.~(\ref{eq:der_tangent})  we get the expression
of the curvature,
\begin{equation}
H^{2}=\left( \frac{\partial^{2}\mathbf{r}}{\partial s^{2}}\right)^{2}.
\end{equation}

The membrane force is deduced from  the functional derivative of the
energy $\delta E/\delta\mathbf{r}$, where $\delta\mathbf{r}$ is a
local small displacement of the vesicle membrane. Due to the
displacement of $\mathbf{r}$ by $\delta\mathbf{r}$,  $ds$ will
undergo variations as well. It is  convenient to introduce a fixed
parametrization (instead of $s$) of the curve, which is denoted by
$\alpha$. $\alpha$ is a parameter that  we can take   to vary from
$0$ to $1$. The correspondence with $s$ is such that $s(\alpha=0)=0$
and $s(\alpha=1)=L$. We then introduce the metric  $g\equiv
\left\vert\partial\mathbf{r}/\partial \alpha \right\vert^{2}$, so
that $ds=\sqrt{g}d\alpha$. We convert the various  terms in the
energy by using now the variable $\alpha$. The curvature assumes the
following expression
\begin{eqnarray}
H^{2} &=& \left(\frac{\partial^2\mathbf{r}}{\partial \alpha^2}\left( \frac{d\alpha}{ds}\right)^{2} +
\frac{\partial\mathbf{r}}{\partial\alpha}\frac{d^{2}a}{ds^{2}}
\right)^{2},\\
&=& \frac{1}{g^{2}}\left(\frac{\partial^{2}\mathbf{r}}{\partial\alpha^{2}}-
\frac{\partial^{2}s}{\partial\alpha^{2}}\mathbf{t}\right)^{2}.
\end{eqnarray}
Writing $\partial^{2}\mathbf{r}/\partial\alpha^{2}$ in terms of the
tangent and the normal vectors, it is  straightforward to show that,
\begin{equation}
\frac{\partial^{2}\mathbf{r}}{\partial\alpha^{2}}=\frac{d^{2}s}{d\alpha^{2}}\mathbf{t}-
gH\mathbf{n},
\end{equation}
This allows to eliminate $s$ from the expression for $H$:
\begin{equation}
H^{2}=\frac{1}{g^{2}}\left(\left( \frac{\partial^{2}\mathbf{r}}{\partial\alpha^{2}}\right)^{2}-
\frac{1}{g}\left(\frac{\partial^{2}\mathbf{r}}{\partial\alpha^{2}}
\frac{\partial\mathbf{r}}{\partial\alpha}\right)^{2}\right).
\label{eq:curvature2}
\end{equation}
\subsection{The curvature force}

Replacing in Eq.~(\ref{eq:energy_c} $H^2$ by the expression given in Eq.~(\ref{eq:curvature2})
we obtain,
\begin{equation}
E_{C}=\frac{\kappa}{2}\int_{0}^{1}\left(\ddot{\mathbf{r}}^{2}
-\frac{1}{g}\left(\ddot{\mathbf{r}}.\dot{\mathbf{r}}\right)^{2}\right)g^{-3/2}d\alpha \,.
\end{equation}
\\
 The
 functional derivative of $E_C$ reads (from classical
 variation results)
\begin{equation}
\frac{\delta E_{C}}{\delta\mathbf{r}}=\frac{\partial e_{C}}{\partial \mathbf{r}}-
\frac{\partial}{\partial \alpha}\frac{\partial e_{C}}{\partial \dot{\mathbf{r}}}+
\frac{\partial^{2}}{\partial \alpha^{2}}\frac{\partial e_{C}}{\partial \ddot{\mathbf{r}}},
\label{eq:fde_c}
\end{equation}
whith $e_{C}=(\kappa/2)\left(\ddot{\mathbf{r}}^{2}
-\frac{1}{g}\left(\ddot{\mathbf{r}}\dot{\mathbf{r}}\right)^{2}\right)g^{-3/2}$,
$\dot{\mathbf{r}}=\partial \mathbf{r}/\partial\alpha$ and
$\ddot{\mathbf{r}}=\partial^{2} \mathbf{r}/\partial\alpha^{2}$.
Since $e_{C}$ does not explicitly depend on $\mathbf{r}$, the first
term on the right hand side of Eq. (\ref{eq:fde_c}) vanishes. The
second term gives,
\begin{eqnarray}
\frac{\partial}{\partial\alpha}\frac{\partial e_{C}}{\partial\dot{\mathbf{r}}} &=&
\kappa\frac{\partial}{\partial\alpha}\left(-\frac{1}{g^{5/2}}\left(
(\ddot{\mathbf{r}}\dot{\mathbf{r}})\ddot{\mathbf{r}}+
\frac{3}{2}(\ddot{\mathbf{r}})^{2}\dot{\mathbf{r}}-
\frac{5}{2g}(\ddot{\mathbf{r}}\dot{\mathbf{r}})^{2}\dot{\mathbf{r}}
\right)\right),\\
&=& -\kappa\frac{\partial}{\partial\alpha}\left(-\frac{\partial^{2}s}{\partial\alpha^{2}}\frac{H}{g}\mathbf{n}+
\frac{3}{2}H^{2}\mathbf{t}\right)
\end{eqnarray}
while the third one becomes,
\begin{eqnarray}
\frac{\partial^{2}}{\partial \alpha^{2}}\frac{\partial e_{C}}{\partial \ddot{\mathbf{r}}} &=&
\kappa
\frac{\partial^{2}}{\partial\alpha^{2}}\left(
\frac{1}{g^{3/2}}\ddot{\mathbf{r}}-\frac{1}{g^{5/2}}(\ddot{\mathbf{r}}\dot{\mathbf{r}})\dot{\mathbf{r}}
\right),\\
&=& \kappa\frac{\partial^{2}}{\partial\alpha^{2}}
\left(
-\frac{H}{\sqrt{g}}\mathbf{n}
\right),\\
&=& \kappa\frac{\partial}{\partial\alpha}
\left(
-\frac{\partial (H\mathbf{n})}{\partial s}+
\frac{\partial^{2}s}{\partial\alpha^{2}}\frac{H}{g}\mathbf{n}
\right).
\end{eqnarray}
Reporting the above results into (\ref{eq:fde_c}),  we obtain the
following expression for the functional derivative
\begin{eqnarray}
\frac{\delta E_{C}}{\delta\mathbf{r}} &=&
\kappa\frac{\partial}{\partial\alpha}
\left(
-\frac{\partial H}{\partial s}\mathbf{n}+
\frac{1}{2}H^{2}\mathbf{t}
\right),\\
&=&-\sqrt{g}\kappa\left(\frac{\partial^{2}H}{\partial s^{2}}+
\frac{1}{2}H^{3}\right)\mathbf{ n},
\end{eqnarray}
Therefore, the membrane curvature force is given by,
\begin{equation}
{\mathbf f_C}=\kappa\left(\frac{\partial^{2}H}{\partial s^{2}}+
\frac{1}{2}H^{3}\right)\mathbf{ n},  \label{eq:f_c}
\end{equation}
where the factor $\sqrt{g}$ disappears from the physical force,
since this one must be defined as ${\mathbf f_C}=-(1/\sqrt{g})
\delta E_C/\delta {\mathbf r}$, as explained at the end of the
appendix.
\subsection{The tension force}

Finally Eq.~(\ref{eq:energy_t}) takes the following form
\begin{equation}
E_{T}=\int_{0}^{1}\zeta(\mathbf{r})\sqrt{g}d\alpha,
\end{equation}
whose functional derivative is,
\begin{equation}
\frac{\delta E_{T}}{\delta\mathbf{r}}=- \frac{\partial}{\partial
\alpha}\frac{\partial e_{T}}{\partial \dot{\mathbf{r}}}
\label{eq:fde_t}
\end{equation}
with $e_{T}=\zeta(\mathbf{r})\sqrt{g}$. Note that  $e_{T}$ depends
neither on $\mathbf{r}$ nor on $\ddot{\mathbf{r}}$. We easily find
\begin{eqnarray}
\frac{\delta E_{T}}{\delta \mathbf{r}}
&=&  -\frac{\partial}{\partial\alpha}\left(
\zeta(\mathbf{r})\frac{\dot{\mathbf{r}}}{\sqrt{g}}\right),\\
&=& -\frac{\partial}{\partial\alpha}\left(
\zeta(\mathbf{r})\mathbf{ t}\right),\\
&=& -\sqrt{g}\frac{\partial}{\partial s}\left(
\zeta(\mathbf{r})\mathbf{ t}\right),\\
&=& -\sqrt{g}\left[\frac{\partial\zeta}{\partial s}\mathbf{ t}-\zeta H \mathbf{ n} \right].
\end{eqnarray}
The membrane force associated with the Lagrange multiplier is then,
\begin{equation}
\mathbf{f}_{T}=-\left[ \zeta H \mathbf{
n}-\frac{\partial\zeta}{\partial s}\mathbf{ t} \right].
\label{eq:f_t}
\end{equation}
By adding Eqs. (\ref{eq:f_c}) and (\ref{eq:f_t}), we obtain the
total membrane force,
\begin{equation}
\mathbf{f}=\left[\kappa\left(\frac{\partial^{2}H}{\partial
s^{2}}+\frac{H^{3}}{2} \right)
-H\zeta\right]\mathbf{n}+\frac{\partial \zeta}{\partial
s}\mathbf{t},
\end{equation}
Let us briefly explain why the force is given by $-{\mathbf
f}=-(1/\sqrt{g}) \delta E_T/\delta {\mathbf r}$ (and not just $
-\delta E_T/\delta {\mathbf r}$). The reason is that what matters is
 a physical displacement of the curve element $ds$ and not
$d\alpha$ (which is a mathematical arbitrary parametrization). If
one performs directly the variation on the integral, one finds
(according to the previous results)
\begin{eqnarray}
\delta E &=&
-\int\left[\left[\kappa\left(\frac{\partial^{2}H}{\partial
s^{2}}+\frac{H^{3}}{2} \right)
-H\zeta\right]\mathbf{n}+\frac{\partial \zeta}{\partial s}\mathbf{t}\right]\sqrt{g}d\alpha\delta\mathbf{r},\\
&=& -\int\left[\left[\kappa\left(\frac{\partial^{2}H}{\partial
s^{2}}+\frac{H^{3}}{2} \right)
-H\zeta\right]\mathbf{n}+\frac{\partial \zeta}{\partial
s}\mathbf{t}\right]ds\delta\mathbf{r}=-\int{\mathbf f} ds
\delta\mathbf{r}
\end{eqnarray}


\begin{thebibliography}{10}

%
%
\bibitem{Kraus96} M. Kraus, W. Wintz, U. Seifert, and R. Lipowsky, Phys. Rev. Lett. 77, 3685
(1996).
\bibitem{seifert99epje} U. Seifert, Eur. Phys. J. B. {\bf 8}, 405 (1999).
%
\bibitem{beaucourt04a}
J. Beaucourt, F. Rioual, T. S\'eon, T. Biben and C. Misbah,
Phys. Rev. E {\bf 69}, 011906 (2004).
%
\bibitem{Gommper} H. Noguchi and G. Gompper, Phys. Rev. Lett. {\bf 93}, 258102
 (2004).
\bibitem{dehaas97}
K.~H. de~Haas, C. Blom, D. van~den Ende, M.~H.~G. Duits and J.
Mellema,
Phys. Rev. E {\bf 56}, 7132 (1997).
%
\bibitem{mader06}
M.-A. Mader, V. Vitkova, M. Abkarian, A. Viallat and T. Podgorski,
Eur. Phys. J. E {\bf 19}, 389 (2006).
%
\bibitem{kantsler06}
V. Kantsler and V. Steinberg, Phys. Re. Lett. {\bf 96}, 036001
(2006).
%
\bibitem{cantat99b}
I. Cantat and C. Misbah,
Phys. Rev. Lett. {\bf 83}, 880 (1999).
%
\bibitem{seifert99}
U. Seifert,
Phys. Rev. Lett. {\bf 83}, 876 (1999).
%
\bibitem{sukumaran01}
S. Sukumaran and U. Seifert,
Phys. Rev. E, {\bf  64}, 011916 (2001)
%
\bibitem{olla}
P. Olla,
J. Phys. A: Math. Gen. {\bf 30}, 317 (1997).
%
\bibitem{abkarian02}
M. Abkarian, C. Lartigue and  A. Viallat,
Phys. Rev. Lett. {\bf 88}, 068103 (2002).
%
\bibitem{beaucourt04b}
J. Beaucourt, T. Biben and C. Misbah,
Europhys. Lett. {\bf 67}, 676 (2004).
%
\bibitem{Leal}
 L.G. Leal,
Ann. Rev. Fluid Mech. {\bf 12}, 435 (1980).


%
\bibitem{Alberts:2001}
B. Alberts, A. Johnson, J. Lewis, M. Raff, K. Roberts and P.
Walter, {\em
  Molecular Biology of the Cell} (Garland Publishing, New York, 2001).
\bibitem{Ladyzhenskaya69}  O. A. Ladyzhenskaya {\em The mathematical theory of
viscous incompresible flow},
  2nd ed. Gordon and Breach, New York, (1969)
%
\bibitem{pozrikidis92}
C. Pozrikidis, {\em Boundary integral and singularity methods for linearized viscous flow}.
(Cambridge University Press, 1992)
%
\bibitem{Helfrich1973}
W. Helfrich, Z. Natureforsch. A {\bf 28c}, 693 (1973).
%
\bibitem{Misbah2006} C. Misbah,  Phys. Rev. Lett. 96, 028104 (2006);  G. Danker, T. Biben ; T. Podgorski, C. Verdier, and C. Misbah, Dynamics and rheology of a dilute suspension of vesicles: higher order theory ,
Phys. Rev. E  (2007), in press.
\bibitem{cantat03} I. Cantat, K. Kassner and C. Misbah,
Eur. Phys. J. E 10, 175 (2003); I. Cantat and C. Misbah, {\em
Transport and structure in biological and chemical systems},
Vol.~532 of {\em Lecture Notes in Physics}, S.~C. M\"uller, J.
Parisi, and W. Zimmermann, eds., (Springer, Heidelberg, 1999), \
pp.\ 93--136.
%
\bibitem{fung} See for example Y.C. Fung, Biomechanics, Springer
(New  york 1990).
%
\bibitem{barthes-biesel}
C. Qu\'eguiner and D. Barth\`es-Biesel, J. Fluid Mech. {\bf 348},
349 (1997).
%
\bibitem{Secomb} T.W. Secomb, R. Skalak, N. Ozkaya, and J.F. Gross,  J. Fluid Mech. {\bf 163}, 405 (1986).

\bibitem{vitkova04}
V. Vitkova, M. Mader and T. Podgorski,
Europhys. Lett., {\bf 68}(3), 398 (2004)
%
%
\bibitem{Joseph:94.1}
J.~Feng, H.~H. Hu and D.~D. Joseph,  J. Fluid Mech. {\bf 277}, 271 (1994).
%
\bibitem{leonhard00}
A. Arend, J. Leonhard, D. Kienle and W. Zimmermann, {\it Cross stream-line migration of bead-spring models
in  nonlinear shear fields} (unpubl.).

\end{thebibliography}
\end{document}